\documentclass[twocolumn,prl,superscriptaddress,showpacs,preprintnumbers,amsmath,amssymb]{revtex4}

\usepackage[ansinew,latin1]{inputenc}
\usepackage{psfrag}
\usepackage{array}
\usepackage{amssymb}
\usepackage{amsmath}
\usepackage{graphicx}

\begin{document}

\title{New constraints on Lorentz invariance violation from the neutron electric dipole moment}
\def\ECTUM{Excellence Cluster `Universe', \TUM}
\def\ETH{ETH Z\"urich, CH-8093 Z\"urich, Switzerland}
\def\GUM{Johannes--Gutenberg--Universit\"at, D--55128 Mainz, Germany}
\def\HNINP{Henryk Niedwodnicza\'nski Institute for Nuclear Physics, 31--342 Cracow, Poland}
\def\ILL{Institut Laue--Langevin, F--38042 Grenoble Cedex, France}
\def\JENA{Department of Neurology, Friedrich--Schiller--University, Jena, Germany}
\def\JINR{JINR, 141980 Dubna, Moscow region, Russia}
\def\JUC{Marian Smoluchowski Institute of Physics, Jagiellonian University, 30--059 Cracow, Poland}
\def\KCGUM{Institut f\"ur Kernchemie, \GUM}
\def\KULEUVEN{Instituut voor Kern-- en Stralingsfysica, Katholieke~Universiteit~Leuven, B--3001 Leuven, Belgium}
\def\LPC{LPC Caen, ENSICAEN, Universit\'e de Caen, CNRS/IN2P3, F--14050 Caen, France}
\def\LPSC{LPSC, Universit\'e Joseph Fourier Grenoble 1, CNRS/IN2P3, Institut National Polytechnique de Grenoble 53, F--38026 Grenoble Cedex, France}
\def\PGUM{Institut f\"ur Physik, \GUM}
\def\PSI{Paul Scherrer Institut (PSI), CH--5232 Villigen-PSI, Switzerland}
\def\RAL{Rutherford Appleton Laboratory, Chilton, Didcot, Oxon OX11 0QX, United Kingdom}
\def\SUSSEX{Department of Physics and Astronomy, University of Sussex, Falmer, Brighton BN1 9QH, United Kingdom}
\def\TUM{Technische Universit\"at M\"unchen, D--85748 Garching, Germany}
\def\UNIFR{University of Fribourg, CH--1700, Fribourg, Switzerland}
\def\UNIVIR{Physics Departement, University of Virginia, Charlottesville, VA, USA}

\author{I. Altarev}      	\affiliation{\TUM}
\author{C.~A.~Baker}     	\affiliation{\RAL}
\author{G.~Ban}          	\affiliation{\LPC}
\author{K.~Bodek}        	\affiliation{\JUC}
\author{M.~Daum}         	\affiliation{\ECTUM} \affiliation{\PSI} \affiliation{\UNIVIR}
\author{M.~Fertl}		\affiliation{\PSI}
\author{B.~Franke}		\affiliation{\ECTUM} \affiliation{\PSI}
\author{P.~Fierlinger}   	\affiliation{\ECTUM}
\author{P.~Geltenbort}   	\affiliation{\ILL}
\author{K.~Green}        	\affiliation{\RAL} \affiliation{\SUSSEX}
\author{M. G. D. van der Grinten} \affiliation{\RAL} \affiliation{\SUSSEX}
\author{P.~G.~Harris}    	\affiliation{\SUSSEX}
\author{R.~Henneck}      	\affiliation{\PSI}
\author{M.~Horras}       	\affiliation{\ECTUM} \affiliation{\PSI}
\author{P.~Iaydjiev}     	\altaffiliation{On leave of absence from INRNE, Sofia, Bulgaria.} \affiliation{\RAL} 
\author{S.~N.~Ivanov}    	\altaffiliation{On leave from PNPI, St Petersburg, Russia.} \affiliation{\RAL}
\author{N.~Khomutov}     	\affiliation{\JINR}
\author{K.~Kirch}        	\affiliation{\PSI} \affiliation{\ETH}
\author{S.~Kistryn}      	\affiliation{\JUC}
\author{A.~Knecht}       	\altaffiliation{Also at University of Z\"urich, Z\"urich, Switzerland. Now at University of Washington, Seattle WA, USA.} \affiliation{\PSI} 
\author{A.~Kozela}       	\affiliation{\HNINP}
\author{F.~Kuchler}     	\affiliation{\ECTUM}
\author{B.~Lauss}        	\affiliation{\PSI}
\author{T.~Lefort}       	\affiliation{\LPC}
\author{Y.~Lemi\`ere}       	\affiliation{\LPC}
\author{A.~Mtchedlishvili}      \affiliation{\PSI}
\author{O.~Naviliat-Cuncic} 	\affiliation{\LPC}
\author{J.~M.~Pendlebury}	\affiliation{\SUSSEX}
\author{G.~Petzoldt}     	\affiliation{\PSI}
\author{E.~Pierre}       	\affiliation{\LPC} \affiliation{\PSI}
\author{F.~M.~Piegsa} 		\affiliation{\ETH}
\author{G.~Pignol}       	\altaffiliation{guillaume.pignol@ph.tum.de} \affiliation{\ECTUM} 
\author{G.~Qu\'em\'ener} 	\affiliation{\LPC} 
\author{D.~Rebreyend}    	\affiliation{\LPSC}
\author{S.~Roccia}       	\altaffiliation{Stephanie.Roccia@fys.kuleuven.be} \affiliation{\KULEUVEN}
\author{P.~Schmidt-Wellenburg}  \affiliation{\PSI}
\author{N.~Severijns}    	\affiliation{\KULEUVEN}
\author{D.~Shiers}       	\affiliation{\SUSSEX}
\author{K.~F.~Smith}		\affiliation{\SUSSEX}
\author{J.~Zejma}        	\affiliation{\JUC}
\author{J.~Zenner}		\affiliation{\PSI} \affiliation{\GUM}
\author{G.~Zsigmond}     	\affiliation{\PSI}

\date{\today}

\begin{abstract}
We propose an original test of Lorentz invariance in the interaction between a particle spin and an electromagnetic field and report on a first measurement using ultracold neutrons.
We used a high sensitivity neutron electric dipole moment (nEDM) spectrometer and searched for a direction dependence of a nEDM signal leading to a modulation of its magnitude at periods of 12 and 24 hours. 
We constrain such a modulation to $d_{12} <  15 \times 10^{-25} \ e\,{\rm cm}$ and $d_{24} <  10  \times 10^{-25} \ e\,{\rm cm}$ at 95~\% C.L.
The result translates into a limit on the energy scale for this type of Lorentz violation effect at the level of ${\cal E}_{LV} > 10^{10}$~GeV.
\end{abstract}


\maketitle

The Standard Model of particle physics (SM) on the one hand and the theory of General Relativity on the other, are the two cornerstones on which our current understanding of the Universe relies. 
Although of seemingly irreconcilable natures, the principle of Lorentz invariance is at the foundation of both theories. 
Unification of these two theories, including a consistent description of the four known interactions, is one of the main challenges of contemporary physics.
Among the many directions being explored, one of the most radical is to abandon spacetime invariance under Lorentz transformations. 
A general framework to parameterize such Lorentz violating (LV) effects has recently been proposed \cite{Colladay}.
It is based on the idea that diluted traces from primordial symmetry breaking can be observed via high precision experiments at low energies. 

Numerous such experiments have been performed over the last century. 
A first category of tests probes the photon sector, with a broad range of techniques from laboratory scale experiments to cosmological observations \cite{Kostelecky2002}. 
A second category deals with particles, including clock comparison experiments, spin polarized torsion pendula and accelerator based experiments. 
The current constraints on Lorentz violating vector and tensor background fields obtained from these experiments have recently been reviewed \cite{Kostelecky2008}. 

In this letter we report on an experimental limit for an interaction between a particle and an electromagnetic field resulting from a fundamental anisotropy of the universe as recently proposed \cite{Bolokhov2006}. 
A nonrelativistic framework will first be developed followed by the description of the experimental procedure and the obtained results. 

Consider a nonrelativistic spin 1/2 particle in the presence of electric and magnetic fields. 
Assuming rotational invariance, the form of the interaction potential is restricted to the simple form $V = - \mu \sigma_i B_i - d \sigma_i E_i$, when considering only the linear terms in the magnetic and electric fields $B_i$ 
and $E_i$. 
Throughout this letter we adopt Einstein's repeated index convention and denote by $\sigma_i$ the Pauli matrices. 
Thus the interaction is described by only two quantities: the magnetic and electric dipole moments $\mu$ and $d$, respectively. 
Hence, allowing for Lorentz violating background vector and tensor fields, in the spirit of \cite{Colladay} and taking into account only linear terms in the electric and magnetic fields, the general form of the interaction potential becomes
\begin{equation}
\label{listeCourses}
V = b_i \sigma_i - d_{ij} \sigma_i E_j - \mu_{ij} \sigma_i B_j. 
\end{equation} 
The first term $b_i$ is sometimes referred to as the cosmic axial field, 
with the dimension of an energy. 
The most stringent limit is $b < 10^{-22}$~eV \cite{Bear2000} but it has been searched for in numerous clock comparison experiments using different particles, including free neutrons \cite{Altarev2009Mod}. 
The next terms $d_{ij}$ and $\mu_{ij}$ in Eq.~(\ref{listeCourses}) have the dimensions of an electric and magnetic dipole moment respectively. 
We will refer to $d_{ij}$ ($\mu_{ij}$) as the cosmic electric (magnetic) dipole tensor. 
They both violate rotation invariance because they define privileged directions in the universe.  

The electric term leads to effects analogous to the electro-optical behavior of anisotropic media. 
If an electric field is applied to a non centrosymmetric medium the latter becomes birefringent for light. 
This is known as the Pockels effect \cite{Pockels}. 
In our case, the vacuum itself is the medium and the particle spin corresponds to the polarization of the light. 

We probed these couplings by observing the spin precession of ultracold neutrons in the presence of a strong electric field and a weak magnetic field, using the RAL/Sussex/ILL spectrometer \cite{Altarev2009} dedicated to the 
search for the neutron 
electric dipole moment \cite{Baker}. 
Under regular experimental conditions, a vertical $B_0 = 1 \, \mu$T magnetic field is applied parallel or antiparallel to a $8.3 \times 10^5$ V/m electric field. 
The sensitivity to the electric term is $10^{4}$ times larger than to the magnetic one due to dimensional considerations. 
Thus, from now on we disregard the anisotropic magnetic moment and focus on the cosmic electric dipole tensor. 

The Ramsey's method of separated oscillating fields was used to measure the Larmor frequency of stored spin-polarized ultracold neutrons. 
Fluctuations of the magnetic field are corrected by means of a spin-polarized $^{199}$Hg vapor as comagnetometer \cite{Green}. 
Both spin-polarized species (ultracold neutrons and mercury atoms) are stored in a cylindrical storage bottle (height $h=~$12~cm, radius $r=~$23.5~cm) during a measurement under vacuum conditions with a duration of 130~s. 
The storage bottle is composed of top and bottom electrodes coated with diamond-like carbon and of an insulating ring coated with deuterated polystyrene \cite{Bodek2008}. 
The homogeneous magnetic field $B_0$ is generated by a coil inside a four-layer mu-metal magnetic shield. 
At the beginning of the precession time, transverse magnetic pulses are applied to flip the polarization of both species by $\pi/2$ onto a plane normal to $B_0$. 
The spin precession of mercury is monitored online by optical means. 
For neutrons a second coherent $\pi/2$ pulse is applied at the end of the precession time. 
The polarization is measured by sequential counting of the number of spin up and down neutrons leaving the storage volume. 

The cosmic electric dipole tensor has in general 9 components and it is convenient to split it into three parts:
\begin{equation}
d_{ij} = d^0 I_{ij} + d_{ij}^{S} + d_{ij}^{A}, 
\end{equation}
where $I$ is the identity matrix, $d^{S}$ is the traceless, symmetric part of the tensor and $d^{A}$ is the antisymmetric part. 
The first term $d^0$ is nothing else than the intrinsic EDM which actually does not violate rotational symmetry. 
The antisymmetric tensor is of rank 2 and dimension 3, thus has $3$ degrees of freedom. 
We define it as the axial vector 
\begin{equation}
d^A_i = \frac{1}{2} \epsilon_{ijk} d^{A}_{jk}, 
\end{equation}
where $\epsilon$ is the completely antisymmetric Levi-Civita tensor. 
Then the antisymmetric part of the interaction potential becomes
\begin{equation}
V^{A} = - d_{ij}^{A} \sigma_i E_j = ( {\bf d}^A \times {\bf E} ) \cdot {\bf \sigma}.
\end{equation}
This potential acts like a magnetic field orthogonal to the electric field.
It is thus orthogonal to the main magnetic field $B_0$ in the apparatus. 
To first order this additional field does not change the Larmor precession frequency and will therefore not be considered any further. 
This is not the case for the five terms arising from the symmetric part of the cosmic EDM tensor which are defined by
\begin{equation}
d_{ij}^S = 
\begin{pmatrix}
d_{XX} & d_{XY} & d_{XZ} \\
d_{XY} & d_{YY} & d_{YZ} \\
d_{XZ} & d_{YZ} & d_{ZZ}
\end{pmatrix},
\end{equation}
with $d_{ZZ}=-(d_{XX}+d_{YY})$. 
These terms contribute to the Larmor precession frequency to first order.
The $Z$ axis is defined as the Earth rotation axis.
While the Earth is rotating together with the vertical quantization axis and the applied electric field, these five contributions would show themselves as an EDM signal in three different ways: 
a steady shift $d_{\rm steady}$ of the value of the EDM, a sidereal modulated part $d_{\rm 24}$ coming from the sidereal modulation of the direction of either the quantization axis or the electric field axis with respect to the static background tensor, and a part modulated at twice the frequency, $d_{\rm 12}$, due to the combined effect of the modulation of both axes. 
Taking into account the intrinsic EDM, we can write these three contributions as:
\begin{eqnarray}
d_{\rm steady} & = & d_0 + \sin^2 \lambda d_{ZZ}  + \frac{ \cos^2 \lambda}{2} (d_{XX} + d_{YY}) \\
d_{\rm 12} & = & \cos^2 \lambda \sqrt{\frac{1}{4}(d_{XX} - d_{YY})^2 + d_{XY}^2}  \\
d_{\rm 24} & = & 2 \cos \lambda \sin \lambda \sqrt{d_{XZ}^2 + d_{YZ}^2}, 
\end{eqnarray}
where $\lambda$ is the latitude of the experimental site. Then 
\begin{equation}
\label{param}
d(t) = d_{\rm steady} + d_{12} \cos(2 \Omega t+\phi_{12}) + d_{24} \cos(\Omega t+\phi_{24})
\end{equation} 
where $\Omega=2 \pi / 23.934$~rad/hour is the sidereal frequency and $\phi_{12}$ and $\phi_{24}$ are phases which depend on the definition of the $X$ and $Y$ axes. 

Following the standard practice in the measurement of the neutron EDM using a comagnetometer, one considers the ratio $R=f_{\rm n}/f_{\rm Hg} \approx 30\,{\rm Hz}/8\,{\rm Hz}$ between the neutron and the mercury precession frequencies. 
In the presence of a homogeneous magnetic field and an electric field, this ratio depends on the direction of the latter according to:
\begin{equation}
R(t) = \left| \frac{\gamma_{\rm n}}{\gamma_{\rm Hg}} + \frac{2E}{h f_{\rm Hg}} \ d(t) \right|
\end{equation}
where $\gamma_{\rm n}$ and $\gamma_{\rm Hg}$ are the gyromagnetic ratios and $d$ is the neutron EDM. 
We neglect a possible contribution from a time-dependent mercury EDM since the Hg nucleus is subject to Schiff screening of the electric field inside the atom \cite{Schiff}. 

\begin{figure}
\includegraphics[width=0.92\linewidth]{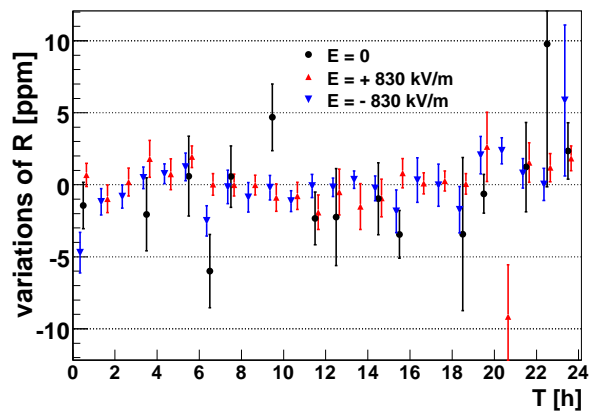}
\caption{Variation of $R$ as a function of time $T$ for electric field up (upwards pointing red triangles) and down (downwards pointing blue triangles) and in the case of a null electric field (black dots). 
For each set of data, the mean value has been substracted.
The data is folded modulo 24~h and then binned. 
} \label{RvsT}
\end{figure}

We studied the time evolution of the correlation between $R$ and the electric field $E$ during 5.6 days in December 2008 at the PF2 ultracold neutron beamline at the Institut Laue-Langevin (ILL), Grenoble. 
An overview of the data is presented in Fig. \ref{RvsT}, where the variation $\Delta R$ of $R$ around the mean value is plotted. 
While the main $B_0$ field was pointing downwards, the electric field was reversed every 2 hours and some additional data were taken without electric field to check for systematic effects. 
The statistical accuracy per cycle of the neutron frequency is given by \cite{Green}:
\begin{equation}
\sigma f_{\rm n} = \frac{1}{2 \pi \sqrt{N} T \alpha_0 e^{-T/T_2}} = 30~\mu\rm{Hz}, 
\end{equation}
where typically $N \approx 4600$ is the number of neutrons per cycle,
$T=130$~s is the precession time and 
$\alpha_0=0.86 \pm 0.01$ is the neutron polarization at the beginning of the
precession time. 
The transverse neutron spin depolarization time $T_2$ depends strongly on the magnetic field homogeneity, whereas the longitudinal depolarization time $T_1=690 \pm 80$~s is attributed to depolarization occurring at wall collisions. 
The field homogeneity can be adjusted by a set of correction coils; the presented data was in fact taken in three different configurations for the currents in these coils. 
For the best magnetic field configuration, the value $T_2 = 400 \pm 38$~s was obtained. 
In addition to the purely statistical error, we expect a fluctuation of the neutron frequency due to a random misalignment of the initial neutron spin after the mercury $\pi/2$ pulse at the level of 20~$\mu$Hz in the worst case. Using only data at zero electric field, we indeed observe a 18~$\mu$Hz non-statistical fluctuation. 
This error was added quadratically to the entire data set. 
The mercury cohabiting magnetometer was performing with a typical accuracy of 0.3~$\mu$Hz or $40$~fT for an averaging time of 130~s. 
Although negligible, the mercury contribution to the individual errors $\sigma R$ was taken into account. 

From a set of $1586$ cycles, one can derive a value for $d_{\rm steady}$ from the difference between $R$ for the two different directions of the electric field:
\begin{equation}
\label{dsteady}
d_{\rm steady} = (-3.4 \pm 2.7_{\rm stat}) \times 10^{-25}~e\,\rm{cm}. 
\end{equation}
Obviously, this $d_{\rm steady}$ term is better 
constrained by the preceding work \cite{Baker} using the same apparatus, i.e. $d_{\rm n} < 2.9\times 10^{-26}~e\,$cm (90 \% C.L.), where statistics has been 
accumulated for several years but where the time evolution has not been studied.
However the fact that the present result was obtained in only about 5 days of data taking shows the high performance of the apparatus. 

\begin{figure}
\includegraphics[width=0.92\linewidth]{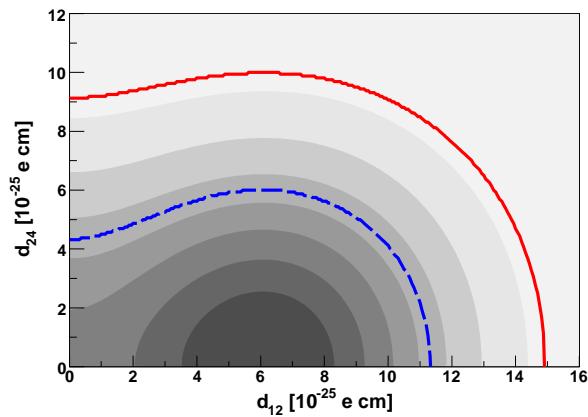}
\caption{Isodensity lines for the posterior probability density function Eq.~(\ref{likelihood}). 
The probability inside the dashed (blue) line is 68 \% and 95 \% inside the solid (red) line. 
} \label{CL2D}
\end{figure}

Further, a Bayesian analysis was applied to the data to search for a time variation $d(t)$, Eq.~(\ref{param}). 
First, the following Chi squared function is established: 
\begin{equation}
\chi^2 (d_{12}, \phi_{12}, d_{24}, \phi_{24}) = 
\sum_{i = 1}^{1586} \left( \frac{ \Delta R_i - \alpha E_i d(t_i)}{\sigma R_i} \right)^2
\end{equation}
where the sum runs over all data cycles and $\alpha = \frac{2}{h f_{\rm Hg}}$. Then the posterior probability density for $d_{12}, d_{24}$ is given by the likelihood function: 
\begin{equation}
\label{likelihood}
L(d_{12}, d_{24}) = \frac{1}{N} \iint \exp(-\chi^2/2) \ d\phi_{12} \ d\phi_{24}
\end{equation}
where $N$ is a normalization coefficient. 
This function is plotted in Fig. \ref{CL2D}, from which we deduce the following bounds: 
\begin{eqnarray}
\label{limits}
\nonumber
d_{12} < & 15 \times 10^{-25} \ e\,{\rm cm} \ & (95 \ \% \ \ \rm{C.L.}) \\
d_{24} < & 10  \times 10^{-25} \ e\,{\rm cm} \ & (95 \ \% \ \ \rm{C.L.})
\end{eqnarray}

This statistical limit could be affected by the main systematic effect namely, a geometrical phase shift of the 
mercury precession frequency proportional to the electric field and the vertical gradient \cite{Pendlebury,Lamoreaux}:
\begin{equation}
\Delta f_{\rm Hg} = \frac{E}{2} \left( \frac{\partial B_0}{\partial z} \right) \left( \frac{\gamma_{\rm Hg}^2 r^2}{c^2} \right) \left[ 1- \left( \frac{\omega_{0}}{\omega_{r}^{\dag}} \right)^2 \right]^{-1}
\label{GeomPhase}
\end{equation}
with $\omega_{0}= |\gamma_{\rm Hg} B_0|$ the Larmor angular frequency and $\omega_{r}^{\dag} = 0.65 \left( v_{xy}/r \right)^2$ the effective 
radial velocity. 
In principle, modulations of the vertical gradient at periods of 12 or 24~h would mimic the signal associated with new physics. 
The magnitude of the vertical gradients can be assessed from the value of $R_0$, the ratio of neutron to mercury precession frequency without 
electric field:
\begin{equation}
R_0 = \left|\frac{f_{\rm n}}{f_{\rm Hg}} \right | = \left| \frac{\gamma_{\rm n}}{\gamma_{\rm Hg}} \left ( 1 - \frac{\partial B_0/\partial z \; \Delta h}{B_0}  \right ) \right |
\label{GravShift}
\end{equation}
which originates from the vertical shift $\Delta h$ of the neutron center of mass with respect to the mercury. 
Due to this gravitational effect, the neutrons and the mercury atoms do not average exactly the same magnetic field in the presence of a vertical gradient. 
By dedicated measurements of $R_0$ and using Eqs.~(\ref{GeomPhase}) and (\ref{GravShift}), it is possible to predict a false electric dipole
moment signal: $d_{\rm false} = (1.2 \pm 0.2) \times 10^{-25}~e\,$cm. 
This shift in $d_{\rm steady}$ is too small to be seen in the data with the given statistics.
The uncertainty in $d_{\rm false}$ has been calculated from the spread in the measured values $R_0$. 
These fluctuations are compatible with statistical fluctuations, in agreement with the previous measurement \cite{Altarev2009Mod}. 
In order to place an upper limit on the contribution of a modulated gradient to our extracted limits $d_{12}$ and $d_{24}$, one can take the uncertainty in $R_0$ as the maximal amount of gradient fluctuations 
according to Eq.~(\ref{GravShift}). 
This then translates via Eq.~(\ref{GeomPhase}) into an upper limit of $2 \times 10^{-26}~e\,$cm as the systematic error in our limits due to gradient modulations. 
Given the current statistical sensitivity, this effect is negligible.

Our result Eq.~(\ref{limits}) can be simply interpreted on the basis of merely dimensional arguments. 
We denote by ${\cal E}_{LV}$ the energy scale associated with a violation of Lorentz invariance. 
It is expected that $d_{12}, d_{24} \approx e \hbar c / {\cal E}_{LV}$. 
This simple argument is supported by more sophisticated arguments in a quantum field theory framework \cite{Bolokhov2006}. 
The limits in Eq. (\ref{limits}) correspond to a lower bound on the energy scale for Lorentz violation effects of $10^{10}$~GeV.
This is far beyond energies accessible at particle colliders ($10^{3}$~GeV) but still below the Grand Unification scale ($10^{16}$~GeV). 
Given that new physics is in general expected to be associated with a large energy scale, the proposed observables $d_{ij}$ are indeed stringent tests of the Lorentz invariance complementary to the search for a 
cosmic axial field. 

A significantly improved sensitivity is expected in the near future with the same experimental installation, which has recently been moved to the Paul Scherrer Institute.
There it will benefit from a more intense ultracold neutron source \cite{Anghel2009}, and upgrades will allow for an even better control of systematic effects \cite{Altarev2009}.

We are grateful to the ILL staff for providing us with excellent running conditions and in particular acknowledge the support of T. Brenner. 
We also benefited from the technical support throughout the collaboration.
This work was partially supported by Polish Ministry of Science and Higher Education, grant No. N202 065436, the Swiss National
Science Foundation, grant No. 200021-126562 and by the DFG cluster of excellence "Origin and
Structure of the Universe".


\end{document}